\documentstyle[prd,aps,epsf,12pt,tighten]{revtex}
\begin{document}

\title{
\hfill{\small {hep-ph/0010043}}\\[2cm]
Lattice regularization for chiral perturbation theory}

\author{Randy Lewis and Pierre-Philippe A. Ouimet}

\address{Department of Physics, University of Regina, Regina, SK, Canada
         S4S 0A2\vspace{3mm}}

\date{October 2000}

\maketitle

\thispagestyle{empty}

\vspace{1cm}

\begin{abstract}
The SU(3) chiral Lagrangian for the lightest octets of mesons and
baryons is constructed on a spacetime lattice.  
The lattice spacing acts as an ultraviolet momentum cutoff which appears 
directly in the Lagrangian so chiral symmetry remains explicit.
As the lattice spacing vanishes, Feynman loop diagrams
typically become divergent due to inverse powers of the lattice spacing, 
and these divergences get absorbed into counterterms such that
the standard results of dimensional regularization are obtained.
One advantage of lattice regularization is that power divergences are
seen explicitly.
In the present work, the octet meson masses, the octet baryon masses
and the $\pi{N}$ sigma term are all computed 
explicitly to one loop order.
\end{abstract}
\newpage

\section{Introduction}

Chiral perturbation theory (ChPT)\cite{chpt} is a low momentum effective 
field theory for QCD.
For the physics of pions, kaons and eta mesons, ChPT organizes the infinite
set of possible interactions into a systematic expansion in inverse powers
of the chiral scale, $\Lambda_\chi$.  Although no precise definition of 
$\Lambda_\chi$ is required, it is understood to be O(1 GeV):
\begin{equation}
\Lambda_\chi \sim m_\rho \sim 4\pi{F_\pi} \sim 4\pi{F_K} \sim 1 {\rm ~GeV}.
\end{equation}
Powers of $4\pi{F}$ appear as natural suppression factors in the calculation
of the loop diagrams in ChPT, and the $\rho$ meson is the lightest hadron which
does not appear explicitly in the ChPT Lagrangian.
The crucial test of ChPT comes from the explicit calculation of observables
to see whether higher order terms in the expansion really give smaller
contributions, and the literature contains many such successful 
examples.\cite{Col}

Since no baryon is light in comparison to $\Lambda_\chi$, it is difficult to
include them into ChPT without destroying the systematic expansion.
Recent work on baryon ChPT is producing interesting
new suggestions\cite{newbchpt}, but we will use the traditional solution,
heavy baryon ChPT (HBChPT)\cite{hbchpt}, which is a double expansion in 
$1/\Lambda_\chi$ and $1/m_{\rm baryon}$.

In the present work, the chiral Lagrangian is constructed on a spacetime
lattice, where the inverse lattice spacing plays the role of a momentum cutoff.
Because the lattice spacing appears directly in the chirally-symmetric
Lagrangian, the chiral properties of calculations are assured.
The lattice-regularized HBChPT Lagrangian is not unique (we will choose an
isotropic hypercubic lattice) but the continuum limit is unique and is
the familiar continuum-defined chiral Lagrangian.
Since observable quantities cannot depend on the regularization and
renormalization prescriptions, any valid lattice implementation must reproduce
the results of dimensional regularization in the limit of vanishing lattice
spacing.

In contrast to dimensional regularization, lattice regularization
shows the power divergences explicitly.
These divergences can also be seen with a nonlattice cutoff, but this
is typically invoked at the time of integration rather than directly in the
Lagrangian, so care is required to be sure that chiral symmetry is not 
violated by the cutoff procedure.
One example of research that relies upon information
about power divergences is found in Ref.~\cite{DonHB}, where
a nonlattice cutoff was used to discuss the convergence of HBChPT and
the scale of baryon substructure.

Lattice QCD simulations are 
always performed with a nonzero lattice spacing,
and they typically use unphysical quark masses as well.
ChPT is regularly employed for the extrapolations to physical 
results.\cite{latextrap}
Lattice-regularized ChPT offers an analytic way to study
discretization errors in these chiral extrapolations.
Although lattice-regularized ChPT does not have a unique Lagrangian,
it does allow a determination of the typical size of discretization
errors as a function of the quark masses and the lattice spacing.
The connection between lattice ChPT and lattice QCD has been
discussed in Ref. \cite{BU} in the limits of strong coupling and large $N_c$.

In section \ref{sec2} of the present work, an SU(3) meson Lagrangian is 
constructed
at leading and next-to-leading chiral order on a spacetime lattice, and
the meson masses and renormalization constants are calculated.
This allows a comparison to the work of Shushpanov and Smilga\cite{ShuS}
who calculated the quadratically-divergent pieces of $F_\pi$, $m_\pi$ and 
wavefunction renormalization in an SU(2) lattice theory.
Section \ref{sec3} of the present work contains the meson-baryon Lagrangian.
A calculation of the baryon masses and the pion-nucleon sigma term is shown to
produce the familiar dimensional regularization results as the lattice spacing
approaches zero.
In section \ref{sec4}, the baryon masses and the pion-nucleon sigma term
are discussed for nonzero lattice spacing, and section \ref{sec5} contains a
concluding discussion.

\section{The meson Lagrangian}\label{sec2}

When the standard SU(3) ChPT Lagrangian of Gasser and Leutwyler\cite{GasL85}
is written in Euclidean spacetime, it takes the following form,
\begin{eqnarray}
{\cal L}_{\rm M} &=& {\cal L}_{\rm M}^{(2)} + {\cal L}_{\rm M}^{(4)}, \\
{\cal L}_{\rm M}^{(2)}  &=& \frac{F^2}{4}{\rm Tr}\{\sum_\mu
   \nabla_\mu{U}^\dagger\nabla_\mu{U}-\chi^\dagger{U}-\chi{U}^\dagger\}, \\
{\cal L}_{\rm M}^{(4)} &=& -L_1\left(\sum_\mu{\rm Tr}\{
   \nabla_\mu{U}^\dagger\nabla_\mu{U}\}\right)^2
   - L_2\sum_{\mu,\nu}{\rm Tr}\{\nabla_\mu{U}^\dagger\nabla_\nu{U}\}
   {\rm Tr}\{\nabla_\mu{U}^\dagger\nabla_\nu{U}\} \nonumber \\
&& - L_3\sum_{\mu,\nu}{\rm Tr}\{\nabla_\mu{U}^\dagger\nabla_\mu{U}
   \nabla_\nu{U}^\dagger\nabla_\nu{U}\}
   + L_4\sum_\mu{\rm Tr}\{\nabla_\mu{U}^\dagger\nabla_\mu{U}\}
   {\rm Tr}\{\chi^\dagger{U}+\chi{U}^\dagger\} \nonumber \\
&& + L_5\sum_\mu{\rm Tr}\{\nabla_\mu{U}^\dagger\nabla_\mu{U}
   (\chi^\dagger{U}+U^\dagger\chi)\}
   - L_6\left({\rm Tr}\{\chi^\dagger{U}+\chi{U}^\dagger\}\right)^2 \nonumber \\
&& - L_7\left({\rm Tr}\{\chi^\dagger{U}-\chi{U}^\dagger\}\right)^2 
   - L_8{\rm Tr}\{\chi^\dagger{U}\chi^\dagger{U}+\chi{U}^\dagger\chi{U}^\dagger
     \} \nonumber \\
&& + iL_9\sum_{\mu,\nu}{\rm Tr}\{F^R_{\mu\nu}\nabla_\mu{U}\nabla_\nu{U}^\dagger
   +F^L_{\mu\nu}\nabla_\mu{U}^\dagger\nabla_\nu{U}\}
   - L_{10}\sum_{\mu,\nu}{\rm Tr}\{U^\dagger{F}^R_{\mu\nu}UF^L_{\mu\nu}\},
\end{eqnarray}
where $U(x)$ is a nonlinear representation of the pseudoscalar meson octet, and
the current quark mass matrix, $\cal M$, enters via 
\begin{equation}
\chi = 2B{\cal M}.
\end{equation}
External vector and axial vector fields appear within the covariant 
derivative,
$\nabla_\mu{U}(x)$, and within the field strengths, $F^L_{\mu\nu}(x)$ and 
$F^R_{\mu\nu}(x)$.

The chiral transformation of $U(x)$ is
\begin{equation}
U(x) \rightarrow g(x)U(x)h(x),
\end{equation} 
where 
$g(x) \in SU_R(3)$ and $h(x) \in SU_L(3)$.
On a spacetime lattice, we introduce the parallel transporters,
$R_\mu(x) \in SU_R(3)$ and $L_\mu(x) \in SU_L(3)$,
which transform as 
\begin{eqnarray}
R_\mu(x) & \rightarrow & g(x)R_\mu(x)g^{\dagger}(x+a_\mu), \\
L_\mu(x) & \rightarrow & h^\dagger(x)L_\mu(x)h(x+a_\mu).
\end{eqnarray}
$a_\mu$ is a Euclidean vector of length $a$ in the $\mu$ direction, and $a$ is
the lattice spacing.
The Lie algebra-valued vector and axial vector fields, $V_\mu(x)$ and 
$A_\mu(x)$, appear in the exponents,
\begin{eqnarray}
L_\mu(x) &=& \exp\{-ia\ell_\mu(x)\}, \\
R_\mu(x) &=& \exp\{-iar_\mu(x)\},
\end{eqnarray}
where $\ell_\mu(x) = V_\mu(x)-A_\mu(x)$ and $r_\mu(x) = V_\mu(x)+A_\mu(x)$.
In ${\cal L}_{\rm M}^{(2)}$, the following covariant derivative is used,
\begin{equation}\label{diffUp}
\nabla_{\mu}^{(+)}U(x) = \frac{1}{a}\left\{R_\mu(x)U(x+a_\mu)L_\mu^\dagger(x) 
            - U(x)\right\},
\end{equation}
but for all other Lagrangian terms (in ${\cal L}_{\rm M}^{(4)}$, 
${\cal L}_{\rm M}^{(6)}$, \ldots) a more symmetric derivative is used:
\begin{equation}\label{diffUpm}
\nabla_{\mu}^{(\pm)}U(x) 
   = \frac{1}{2a}\left\{R_\mu(x)U(x+a_\mu)L_\mu^\dagger(x) 
                      - R_\mu^\dagger(x-a_\mu)U(x-a_\mu)L_\mu(x-a_\mu)\right\}.
\end{equation}
Use of $\nabla_{\mu}^{(\pm)}$ at leading order would produce extraneous
poles in the meson propagator (``doublers'') and use of $\nabla_{\mu}^{(+)}$
in higher order Lagrangian terms can violate parity.

The lattice ChPT action is simply obtained by summing the Lagrangian over
all spacetime lattice sites,
\begin{equation}
S_{\rm M}[U,V_\mu,A_\mu] = a^4\sum_x{\cal L}_{\rm M}(x).
\end{equation}
The propagators and vertices required for perturbative calculations
can be extracted from this action, but if the path integral 
formalism is used, one must pay
particular attention to extra meson interactions which get generated by the
integration measure, 
\begin{equation}
DU = e^{-S_{\rm meas}[\pi]}\prod_{a=1}^8{\rm d}\pi^a,
\end{equation}
The definition $U(x)=\exp\{-i\lambda^a\pi^a(x)/F\}$ has been employed;
$\lambda^a$ is a Gell-Mann matrix and $\pi^a(x)$ is a pseudoscalar meson field.
The ``effective action'' from the measure is\cite{Rothe}
\begin{eqnarray}
S_{\rm meas}[\pi] &=& 
        -\frac{1}{2}\sum_x{\rm Tr}\ln\left\{\frac{2(1-\cos\Phi(x))}
                  {\Phi^2(x)}\right\}, \\
\Phi(x) &=& \frac{2}{F}\sum_{a=1}^8t^a\pi^a(x),
\end{eqnarray}
where $t^a_{bc}=-if_{abc}$ and $f_{abc}$ are the
structure constants defined by $[\lambda^a,\lambda^b]=2if^{abc}\lambda^{c}$.
These measure contributions also exist in the continuum theory, although
they happen to vanish when dimensional regularization is used.

Neglecting isospin violation and using $m_l$ to denote the up and down
quark masses,
the lowest order pion, kaon and eta two point functions are
\begin{equation}\label{neginvprop}
\Gamma_{MM} = 
-\left\{x_M^2+\frac{4}{a^2}\sum_\mu\sin^2\left(\frac{aq_\mu}{2}\right)\right\},
\end{equation}
where 
\begin{eqnarray}
x_\pi &=& \sqrt{2Bm_l}, \\
x_K &=& \sqrt{B(m_l+m_s)}, \\
x_\eta &=& \sqrt{\frac{2}{3}B(m_l+2m_s)}.
\end{eqnarray}
The meson masses are obtained from the zero of Eq.~(\ref{neginvprop}), 
corresponding to the pole in the propagator,
\begin{equation}\label{physmass}
m_M = \frac{2}{a}{\rm arcsinh}\left(\frac{ax_M}{2}\right).
\end{equation}
Notice the existence of a Gell-Mann--Okubo relation,
\begin{equation}\label{mesonGMO}
3{\rm sinh}^2\left(\frac{am_\eta}{2}\right)
= 4{\rm sinh}^2\left(\frac{am_K}{2}\right)
- {\rm sinh}^2\left(\frac{am_\pi}{2}\right),
\end{equation}
which reproduces the conventional relation as $a \rightarrow 0$.

\begin{figure}[tbh]
\epsfxsize=380pt \epsfbox[30 320 498 450]{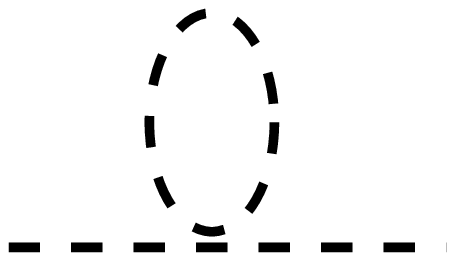}
\caption{The only meson loop topology which contributes to a meson mass
         at one-loop order in ChPT.
         }\label{fig:mloop}
\end{figure}

At next-to-leading order, the meson masses receive 
tree-level contributions from ${\cal L}_{\rm M}^{(4)}$ and from 
$S_{\rm meas}[\pi]$ as well as
loop diagrams from the interactions of
${\cal L}_{\rm M}^{(2)}$.  The only loop topology at this order is shown
in Fig.~\ref{fig:mloop} and,
for example, a charged kaon loop makes the following addition to the
pion two-point function,
\begin{eqnarray}
\Delta\Gamma_{\pi\pi} &=& \frac{1}{3a^2F^2}
\int_{-\pi/a}^{\pi/a}\frac{{\rm d}^4p}{(2\pi)^4}
\left\{x_K^2+\frac{4}{a^2}\sum_\nu\sin^2\left(\frac{ap_\nu}{2}\right)
\right\}^{-1} \nonumber \\
&\times& \left(\frac{a^2}{2}(x_\pi^2+x_K^2)
+\sum_\mu\{5-4\cos(aq_\mu)-4\cos(ap_\mu)+3\cos(aq_\mu-ap_\mu)\}\right).
\label{sampleloop}
\end{eqnarray}
Notice that only momenta within the first Brillouin zone can appear in the
integral, since physics at shorter distances cannot be resolved on the
spacetime lattice.  The integral is thus finite, and only diverges as
$a \rightarrow 0$.
It is convenient to rewrite the propagator within Eq.~(\ref{sampleloop})
as the integral of an exponential, using 
$1/D = \int_0^\infty{\rm d}x\,\exp(-xD)$, which leads to 
\begin{eqnarray}
\Delta\Gamma_{\pi\pi} &=& 
\frac{1}{6a^4F^2}\left\{1+\frac{a^2x_\pi^2}{2}W_4(a^2x_K^2)\right\} \nonumber\\
&+& \frac{1}{4a^4F^2}\left\{1+\frac{4}{3}\left(1-\frac{3}{8}a^2x_K^2\right)
W_4(a^2x_K^2)\right\}\sum_\mu\sin^2\left(\frac{aq_\mu}{2}\right),
\end{eqnarray}
where
\begin{equation}\label{defW}
W_n(\epsilon^2) \equiv \int_0^\infty{\rm d}x\,I_0^n(x)\exp\left\{-x\left(n+
\frac{\epsilon^2}{2}\right)\right\}
\end{equation}
and $I_0(x)$ is a Bessel function,
\begin{equation}
I_0(x) = \int_{-\pi}^\pi\frac{{\rm d}\theta}{2\pi}\,\exp(x\cos\theta).
\end{equation}
By including all of the one-loop diagrams and ${\cal L}_{\rm M}^{(4)}$
tree-level pieces,
the complete two point functions to next-to-leading order are found to be
\begin{equation}
\Gamma_{MM} = -\frac{1}{Z_M^{(+)}Z_M^{(\pm)}}\left\{X_M^2+\frac{4Z_M^{(\pm)}}
              {a^2}\sum_\mu\sin^2\left(\frac{aq_\mu}{2}\right)
            +  \left(\frac{1-Z_M^{(\pm)}}{a^2}\right)
              \sum_\mu\sin^2\left(aq_\mu\right)\right\}
\end{equation}
with
\begin{eqnarray}
Z_\pi^{(+)} &=& 1 + \frac{7}{24a^2F^2} 
        + \frac{1}{3a^2F^2}\left(1-\frac{3}{16}a^2x_\pi^2\right)W_4(a^2x_\pi^2)
        \nonumber \\
     && + \frac{1}{6a^2F^2}\left(1-\frac{3}{8}a^2x_K^2\right)W_4(a^2x_K^2)
        - \frac{x_\eta^2}{48F^2}W_4(a^2x_\eta^2), \\
Z_\pi^{(\pm)} &=& 1 - \frac{8}{F^2}(x_\pi^2+2x_K^2)L_4 
                    - \frac{8}{F^2}x_\pi^2L_5, \\
Z_K^{(+)} &=& 1 + \frac{7}{24a^2F^2} 
        + \frac{1}{8a^2F^2}\left(1-\frac{3}{8}a^2x_\pi^2\right)W_4(a^2x_\pi^2)
        \nonumber \\
     && + \frac{1}{4a^2F^2}\left(1-\frac{3}{8}a^2x_K^2\right)W_4(a^2x_K^2)
  + \frac{1}{8a^2F^2}\left(1-\frac{1}{24}a^2x_\eta^2\right)W_4(a^2x_\eta^2),\\
Z_K^{(\pm)} &=& 1 - \frac{8}{F^2}(x_\pi^2+2x_K^2)L_4 - \frac{8}{F^2}x_K^2L_5,\\
Z_\eta^{(+)} &=& 1 + \frac{7}{24a^2F^2} 
        - \frac{x_\pi^2}{16F^2}W_4(a^2x_\pi^2) \nonumber \\
     && + \frac{1}{2a^2F^2}\left(1-\frac{1}{24}a^2x_K^2\right)W_4(a^2x_K^2)
        - \frac{x_\eta^2}{16F^2}W_4(a^2x_\eta^2), \\
Z_\eta^{(\pm)} &=& 1 - \frac{8}{F^2}(x_\pi^2+2x_K^2)L_4 
                     - \frac{8}{F^2}x_\eta^2L_5, \\
X_\pi^2 &=& x_\pi^2 - \frac{8}{F^2}x_\pi^2(x_\pi^2+2x_K^2)(L_4-2L_6)
        - \frac{8}{F^2}x_\pi^4(L_5-2L_8)
        \nonumber \\
     && + \frac{7x_\pi^2}{24a^2F^2}
        + \frac{x_\pi^2}{4a^2F^2}\left\{W_4(a^2x_\pi^2)-\frac{1}{3}W_4(a^2
          x_\eta^2)\right\} \nonumber \\
     && - \frac{x_\pi^4}{16F^2}W_4(a^2x_\pi^2)
        - \frac{x_\pi^2x_K^2}{16F^2}W_4(a^2x_K^2)
        - \frac{x_\pi^2x_\eta^2}{48F^2}W_4(a^2x_\eta^2), \\
X_K^2 &=& x_K^2 - \frac{8}{F^2}x_K^2(x_\pi^2+2x_K^2)(L_4-2L_6)
        - \frac{8}{F^2}x_K^4(L_5-2L_8)
        \nonumber \\
     && + \frac{7x_K^2}{24a^2F^2}
        + \frac{x_K^2}{6a^2F^2}W_4(a^2x_\eta^2)
        \nonumber \\
     && - \frac{3x_\pi^2x_K^2}{64F^2}W_4(a^2x_\pi^2)
        - \frac{3x_K^4}{32F^2}W_4(a^2x_K^2)
        - \frac{x_K^2x_\eta^2}{192F^2}W_4(a^2x_\eta^2), \\
X_\eta^2 &=& x_\eta^2 - \frac{8}{F^2}x_\eta^2(x_\pi^2+2x_K^2)(L_4-2L_6)
        - \frac{8}{F^2}x_\eta^4(L_5-2L_8) + \frac{128}{9F^2}(x_K^2-x_\pi^2)^2
          (3L_7+L_8) \nonumber \\
     && + \frac{7x_\eta^2}{24a^2F^2}
        - \frac{x_\pi^2}{4a^2F^2}W_4(a^2x_\pi^2)
        + \frac{(x_\pi^2+3x_\eta^2)}{6a^2F^2}W_4(a^2x_K^2)
        + \frac{(7x_\pi^2-16x_K^2)}{36a^2F^2}W_4(a^2x_\eta^2)
        \nonumber \\
     && - \frac{x_\pi^2x_\eta^2}{16F^2}W_4(a^2x_\pi^2)
        - \frac{x_K^2x_\eta^2}{48F^2}W_4(a^2x_K^2)
        - \frac{x_\eta^4}{16F^2}W_4(a^2x_\eta^2).
\end{eqnarray}
The physical meson masses are obtained by using $X_M$ instead of $x_M$ in
Eq.~(\ref{physmass}).
Notice that the pseudoscalar mesons are exactly massless in 
the chiral limit ($m_l=m_s=0$), indicating that the theory does indeed
have exact chiral symmetry even for nonzero lattice spacing.
This feature has been emphasized in the SU(2) case by the authors of
Ref.~\cite{ShuS}.

The lattice regularized theory gives finite predictions for all observables
at nonzero lattice spacing.  The chiral expansion is in inverse powers of
$\Lambda_\chi$.  If one chooses to extrapolate to the limit of vanishing 
lattice spacing, then divergences appear in the loop
contributions and they must be absorbed into renormalized values of the
Lagrangian parameters.  With the information in Appendix~\ref{app:W}, it is a
simple matter to show that the continuum limit of the lattice theory gives
precisely the masses that are familiar from dimensional 
regularization.\cite{GasL85}  Moreover, the logarithmic dependences of the
counterterms are found to be
\begin{eqnarray}
L_4^r(1/a_2)-2L_6^r(1/a_2)-\{L_4^r(1/a_1)-2L_6^r(1/a_1)\} &=& 
-\frac{1}{36(4\pi)^2}\ln\left(\frac{a_2}{a_1}\right), \\
L_5^r(1/a_2)-2L_8^r(1/a_2)-\{L_5^r(1/a_1)-2L_8^r(1/a_1)\} &=& 
\frac{1}{6(4\pi)^2}\ln\left(\frac{a_2}{a_1}\right), \\
3L_7^r(1/a_2)+L_8^r(1/a_2)-\{3L_7^r(1/a_1)+L_8^r(1/a_1)\} &=& 
\frac{5}{48(4\pi)^2}\ln\left(\frac{a_2}{a_1}\right),
\end{eqnarray}
for sufficiently small lattice spacings $a_1$ and $a_2$.  This is precisely
the scale dependence that is known from Ref.~\cite{GasL85}, as required,
since observables cannot depend on the regularization prescription.

\section{The meson-baryon Lagrangian}\label{sec3}

The HBChPT Lagrangian is organized as
a systematic expansion in the inverse baryon mass as well as the inverse
chiral scale, $\Lambda_\chi$.  This is accomplished by writing the Lagrangian
in terms of a heavy baryon field, $B_v(x)$, instead of the relativistic
field, $B(x)$, as follows,
\begin{equation}
B_v(x) = \exp(im_{\rm HB}v\cdot{x})\frac{1}{2}(1+v\!\!\!/)B(x),
\end{equation}
where the mass parameter $m_{\rm HB}$ is chosen to cancel, or nearly cancel,
the octet baryon masses.  The first few orders
in the double expansion of HBChPT are well known\cite{hbchpt},
and in Euclidean spacetime one finds
\begin{eqnarray}
{\mathcal{L}}_{\rm MB} &=& 
        {\mathcal{L}}_{\rm MB}^{(0)} + {\mathcal{L}}_{\rm MB}^{(1)}
      + {\mathcal{L}}_{\rm MB}^{(2)} + {\mathcal{L}}_{\rm MB}^{(3)}
      + {\rm ~higher~order}, \\
{\mathcal{L}}_{\rm MB}^{(0)} &=& (m_0-m_{\rm HB}){\rm Tr}\left(\bar{B}_v
 B_v\right), \\
{\mathcal{L}}_{\rm MB}^{(1)} &=& \sum_\mu \bigg[{\rm Tr}\left(\bar{B}_vv_\mu
 D_\mu B_v\right)+{\mathcal{D}}{\rm Tr}\left(\bar{B}_vS_\mu\{u_\mu,B_v\}\right)
+{\mathcal{F}}{\rm Tr}\left(\bar{B}_vS_\mu[u_\mu,B_v]\right)\bigg], \\
{\mathcal{L}}_{\rm MB}^{(2)} &=& 
\frac{1}{2m_0}{\rm Tr}\left(\bar{B}_v(v\cdot{D}v\cdot{D}-D^2)B_v\right)
-b_{\mathcal{D}}{\rm Tr}\left(\bar{B}_v\{\chi_+,B_v\}\right)
-b_{\mathcal{F}}{\rm Tr}\left(\bar{B}_v[\chi_+,B_v]\right) \nonumber \\
&&-b_0{\rm Tr}\left(\bar{B}_vB_v\right){\rm Tr}\left(\chi_+\right) +\ldots, \\
{\mathcal{L}}_{\rm MB}^{(3)} &=& \ldots,
\end{eqnarray}
where the omitted terms do not contribute to the present work.
$m_0$ is the leading contribution to the octet baryon mass that would
appear in a relativistic Lagrangian, 
$\mathcal{D}$ and $\mathcal{F}$ are the two axial couplings, and
$S_\mu=\frac{i}{2}\gamma_5\sum_\nu\sigma_{\mu\nu}v_\nu$
is the Pauli-Lubanski spin vector.
The matrix $B_v$ denotes the baryon octet,
\begin{eqnarray}
B_v  =  \left(
\matrix  { {1\over \sqrt 2} \Sigma^0_v + {1 \over \sqrt 6} \Lambda_v
&\Sigma^+_v &  p_v \nonumber \\
\Sigma^-_v
    & -{1\over \sqrt 2} \Sigma^0_v + {1 \over \sqrt 6} \Lambda_v & n_v
    \nonumber \\
\Xi^-_v
        &       \Xi^0_v &- {2 \over \sqrt 6} \Lambda_v \nonumber \\}
\!\!\!\!\!\!\!\!\!\!\!\!\!\!\!\!\! \right),
\end{eqnarray}
while the pseudoscalar mesons appear within $U=\xi^2$ and 
$\chi_+ = \xi^\dagger\chi\xi^\dagger + \xi\chi^\dagger\xi$.
The hadron fields transform under local chiral transformations as
\begin{eqnarray}
B_v(x) &\rightarrow& o(x)B_v(x)o^\dagger(x), \\
\xi(x) &\rightarrow& g(x)\xi(x)o^\dagger(x) = o(x)\xi(x)h(x).
\end{eqnarray}

It is convenient to choose $v=(0,0,0,1)$, and to use a nearest-neighbour
covariant derivative in the time direction,
\begin{eqnarray}
aD_4B_v(x)&=&B_v(x) \nonumber \\
&-& \frac{1}{4}\xi^\dagger(x)R^\dagger_4(x-a_4)\xi(x-a_4)B_v(x-a_4)
    \xi^\dagger(x-a_4)R_4(x-a_4)\xi(x) \nonumber \\
&-& \frac{1}{4}\xi(x)L^\dagger_4(x-a_4)\xi^\dagger(x-a_4)B_v(x-a_4)
    \xi(x-a_4)L_4(x-a_4)\xi^\dagger(x) \nonumber \\
&-& \frac{1}{4}\xi^\dagger(x)R^\dagger_4(x-a_4)\xi(x-a_4)B_v(x-a_4)
    \xi(x-a_4)L_4(x-a_4)\xi^\dagger(x) \nonumber \\
&-& \frac{1}{4}\xi(x)L^\dagger_4(x-a_4)\xi^\dagger(x-a_4)B_v(x-a_4)
    \xi^\dagger(x-a_4)R_4(x-a_4)\xi(x).
\end{eqnarray}
Spatial covariant derivatives appear in 
${\cal L}_{\rm MB}^{(2)} + {\cal L}_{\rm MB}^{(3)} + \ldots$, and for
these a symmetric definition is used (to conserve parity):
\begin{eqnarray}
aD_jB_v(x)&=&
    \frac{1}{8}\xi(x)L_j(x)\xi^\dagger(x+a_j)B_v(x+a_j)
    \xi(x+a_j)L_j^\dagger(x)\xi^\dagger(x) \nonumber \\
&+& \frac{1}{8}\xi^\dagger(x)R_j(x)\xi(x+a_j)B_v(x+a_j)
    \xi(x+a_j)L_j^\dagger(x)\xi^\dagger(x) \nonumber \\
&+& \frac{1}{8}\xi(x)L_j(x)\xi^\dagger(x+a_j)B_v(x+a_j)
    \xi^\dagger(x+a_j)R_j^\dagger(x)\xi(x) \nonumber \\
&+& \frac{1}{8}\xi^\dagger(x)R_j(x)\xi(x+a_j)B_v(x+a_j)
    \xi^\dagger(x+a_j)R_j^\dagger(x)\xi(x) \nonumber \\
&-& \frac{1}{8}\xi^\dagger(x)R^\dagger_j(x-a_j)\xi(x-a_j)B_v(x-a_j)
    \xi^\dagger(x-a_j)R_j(x-a_j)\xi(x) \nonumber \\
&-& \frac{1}{8}\xi(x)L^\dagger_j(x-a_j)\xi^\dagger(x-a_j)B_v(x-a_j)
    \xi(x-a_j)L_j(x-a_j)\xi^\dagger(x) \nonumber \\
&-& \frac{1}{8}\xi^\dagger(x)R^\dagger_j(x-a_j)\xi(x-a_j)B_v(x-a_j)
    \xi(x-a_j)L_j(x-a_j)\xi^\dagger(x) \nonumber \\
&-& \frac{1}{8}\xi(x)L^\dagger_j(x-a_j)\xi^\dagger(x-a_j)B_v(x-a_j)
    \xi^\dagger(x-a_j)R_j(x-a_j)\xi(x).
\end{eqnarray}
One also defines
\begin{equation}
u_\mu(x) = \frac{i}{2}\xi^\dagger(x)\nabla_\mu^{(\pm)}U(x)\xi^\dagger(x)
         - \frac{i}{2}\xi(x)\nabla_\mu^{(\pm)}U^\dagger(x)\xi(x),
\end{equation}
where $\nabla_\mu^{(\pm)}U(x)$ is given by Eq.~(\ref{diffUpm}). 

{}From ${\mathcal{L}}_{\rm MB}^{(0)} + {\mathcal{L}}_{\rm MB}^{(1)}$,
the lowest order baryon two-point function is
\begin{equation}
\Gamma_{BB} = m_{\rm HB} - m_0
                  -\frac{i}{a}\left\{\sin(aq_4)
                  -2i\sin^2\left(\frac{aq_4}{2}\right)\right\},
\end{equation}
which has a unique zero in the first Brillouin zone, occurring at
\begin{equation}\label{LObarymass}
E \equiv -ip_4 = \frac{1}{a}\ln\{1+a(m_0-m_{\rm HB})\}.
\end{equation}
The physical baryon mass is then $m_{\rm HB}+E = m_0 + O(a)$.
Typically the parameter $m_{\rm HB}$ is chosen to equal $m_0$,
and then $E=0$ at this order in lattice-regularized HBChPT.

The contribution of ${\mathcal{L}}_{\rm MB}^{(2)}$ to the two-point function is
purely tree-level and the contribution at the order 
of ${\mathcal{L}}_{\rm MB}^{(3)}$ is
purely from loop diagrams.  The two topologies for loop diagrams are shown in 
Fig.~\ref{fig:bloops}.  The diagram with no internal baryon propagator
involves the same functions that were used in the previous section for meson
masses.  For example, the contribution of the charged pion loop to the 
proton two-point function is
\begin{eqnarray}
\Delta\Gamma_{pp}^{(a)} &=& \frac{i}{2aF^2}
\int_{-\pi/a}^{\pi/a}\frac{{\rm d}^4p}{(2\pi)^4}
\left\{x_\pi^2+\frac{4}{a^2}\sum_\nu\sin^2\left(\frac{ap_\nu}{2}\right)
\right\}^{-1} \nonumber \\
&\times& \left\{\sin(aq_4)-\sin(aq_4-ap_4)
         -2i\sin^2\left(\frac{aq_4}{2}\right)
         +2i\sin^2\left(\frac{aq_4-ap_4}{2}\right)\right\}, \\
&=& -\frac{1}{16a^3F^2}\{\cos(aq_4)-i\sin(aq_4)\}
    \left\{1-\frac{1}{2}a^2x_\pi^2W_4(a^2x_\pi^2)\right\}.
\end{eqnarray}
Because the other diagram in Fig.~\ref{fig:bloops} has an
internal baryon propagator, its evaluation is somewhat more involved.
The calculation is outlined in Appendix~\ref{app:loop}.

As anticipated by Eq.~(\ref{LObarymass}), the final results for 
the baryon masses are
\begin{equation}\label{barymass}
m_B = m_{\rm HB} + \frac{1}{a}\ln(1+aX_B)
\end{equation}
where
\begin{eqnarray}
X_B &=& m_0 - m_{\rm HB} - 2(2x_K^2+x_\pi^2)b_0
     + \gamma_B^{\cal D}b_{\cal D} + \gamma_B^{\cal F}b_{\cal F}
   - \frac{1}{3a^3F^2} \nonumber \\
     &-& \frac{1}{16a^3F^2}\sum_{i=\pi,K,\eta}\alpha_B^i\overline{Y}(a^2x_i^2)
      +  \frac{1}{16aF^2}\sum_{i=\pi,K,\eta}x_i^2\beta_B^iW_4(a^2x_i^2). 
\end{eqnarray}
$\overline{Y}(\epsilon^2) \equiv Y_3(\epsilon^2) + Y_4(\epsilon^2)$ is 
defined by Eq.~(\ref{Yn}), and the coefficients $\alpha_B^i$ and $\beta_B^i$
are given in Table \ref{tab:coeffs}.

As $a\rightarrow 0$ these baryon masses must be identical to the results
of dimensional regularization.  Unlike dimensional regularization, the
results of Eq.~(\ref{barymass}) contain cubic and linear
divergences as $a\rightarrow 0$, but these can be absorbed into renormalized
parameters as follows,
\begin{eqnarray}
m_0^r &=& m_0 - \frac{1}{3a^3F^2} - \frac{(5{\cal D}^2+9{\cal F}^2)}
          {16a^3F^2}\overline{Y}(0), \\
b_0^r &=& b_0 + \frac{(13{\cal D}^2+9{\cal F}^2)}{192aF^2}\overline{Y}^\prime(0)
              - \frac{11}{576aF^2}W_4(0), \\
b_{\cal D}^r &=& b_{\cal D}
              - \frac{3({\cal D}^2-3{\cal F}^2)}{128aF^2}\overline{Y}^\prime(0) 
              - \frac{5}{384aF^2}W_4(0), \\
b_{\cal F}^r &=& b_{\cal F} + \frac{5{\cal DF}}{64aF^2}\overline{Y}^\prime(0),
\end{eqnarray}
where a prime denotes differentation with respect to the argument.
It is convenient to choose $m_{\rm HB} = m_0^r$.
With reference to Appendix~\ref{app:W}, the $a\rightarrow 0$ limit is
easily obtained, and is identical to the known dimensional regularization
results\cite{BKM} as required,
\begin{equation}
m_B \rightarrow
m_0^r - 2(2m_K^2+m_\pi^2)b_0^r + \gamma_B^{\cal D}b_{\cal D}^r
+ \gamma_B^{\cal F}b_{\cal F}^r - \frac{(\alpha_B^{\pi}m_\pi^3
+\alpha_B^Km_K^3+\alpha_B^{\eta}m_\eta^3)}{24\pi{F}^2}.
\end{equation}

To conclude this section, consider the pion-nucleon sigma term defined at
zero momentum transfer via the Feynman-Hellman theorem,
\begin{equation}
\sigma_{\pi{N}} = \hat{m}\frac{\partial{m}_N}{\partial\hat{m}},
\end{equation}
where $\hat{m} = (m_u+m_d)/2$.  With lattice regularization and exact isospin
symmetry, the relation becomes
\begin{equation}
\sigma_{\pi{N}} = x_\pi^2\left(\frac{\partial~~~~~}{\partial(x_\pi^2)}
                + \frac{1}{2}\frac{\partial~~~~~}{\partial(x_K^2)}\right)X_N,
\end{equation}
leading to the following explicit expression,
\begin{eqnarray}
\sigma_{\pi N} &=& -2x_\pi^2(2b_0+b_{\cal D}+b_{\cal F}) \nonumber \\
 &+&\frac{3x_\pi^2}{64aF^2}\{W_4(a^2x_\pi^2)+a^2x_\pi^2W_4^\prime(a^2x_\pi^2)
     +W_4(a^2x_K^2)+a^2x_K^2W_4^\prime(a^2x_K^2)\} \nonumber \\
  &+&\frac{5x_\pi^2}{576aF^2}\{W_4(a^2x_\eta^2)+a^2x_\eta^2W_4^\prime
     (a^2x_\eta^2)\} \nonumber \\
  &-&\frac{9x_\pi^2}{64aF^2}({\cal D}+{\cal F})^2\overline{Y}^\prime
     (a^2x_\pi^2) \nonumber \\
  &-&\frac{x_\pi^2}{64aF^2}(5{\cal D}^2-6{\cal DF}+9{\cal F}^2)
     \overline{Y}^\prime(a^2x_K^2) \nonumber \\
  &-&\frac{x_\pi^2}{192aF^2}({\cal D}-3{\cal F})^2
     \overline{Y}^\prime(a^2x_\eta^2).
     \label{sigma}
\end{eqnarray}
Using Appendix~\ref{app:W}, the $a\rightarrow 0$ limit is found to be
\begin{eqnarray}
\sigma_{\pi N} &\rightarrow& -2m_\pi^2(2b_0^r + b_{\cal D}^r + b_{\cal F}^r)
- \frac{9m_\pi^3}{64 \pi F^2}({\cal D}+{\cal F})^2
- \frac{m_\pi^2m_K}{64 \pi F^2}(5{\cal D}^2-6{\cal DF}+9{\cal F}^2) \nonumber\\
& & - \frac{m_\pi^2m_\eta}{192 \pi F^2}({\cal D}-3{\cal F})^2,
\end{eqnarray}
as has been obtained from dimensional regularization\cite{BKM}.

\section{The baryon masses at nonzero lattice spacing}\label{sec4}

In the previous sections of this work, it has been shown that expressions
for the meson masses, the baryon masses and the $\pi N$ sigma term are
the same in both dimensional regularization and lattice regularization
in the limit of vanishing lattice spacing.
Different expressions are obtained when $a \neq 0$.

Consider first a lattice spacing that satisfies $\pi/a>\Lambda_\chi$.
Most lattice QCD simulations are performed with lattice
spacings that satisfy this criterion, but with quark masses that are
larger than the physical values.  If lattice QCD computations
are first extrapolated to the continuum, then the chiral extrapolations
(i.e. the extrapolations of observables from the simulated quark masses
to the physical quark masses)
can use the continuum ChPT Lagrangian.  Without the initial extrapolation
to the continuum, lattice QCD data should obey a lattice ChPT Lagrangian
instead of the continuum one.
In practice, the $a \neq 0$ effects of
lattice-regularized ChPT should be $O(pa/\pi)<O(p/\Lambda_\chi)$, where 
$p\sim m_\pi$ is a typical momentum.
This can now be tested explicitly for the observables under discussion.

For completeness, we also consider a coarser lattice satisfying
$\pi/a<\Lambda_\chi$.
In this case, Eq.~(\ref{barymass}) can be expanded 
in powers of $x_i/(4\pi{F})$, $x_i/m_0$, $(\pi/a)/(4\pi{F})$ and $(\pi/a)/m_0$
as follows,
\begin{eqnarray}\label{mBexpansion}
m_B &=& m_{\rm HB} + X_B + {\rm ~higher~order} \nonumber \\
    &=& m_B^{(0)} + m_B^{(1)} + m_B^{(2)} + m_B^{(3)} + {\rm ~higher~order}, \\
m_B^{(0)} &=& m_0, \\
m_B^{(1)} &=& 0, \\
m_B^{(2)} &=& -2(2x_K^2+x_\pi^2)b_0 + \gamma_B^{\cal D}b_{\cal D}
              + \gamma_B^{\cal F}b_{\cal F}, \label{mB2} \\
m_B^{(3)} &=& -\frac{a}{2}\left(m_B^{(2)}\right)^2 
              - \frac{1}{3a^3F^2} - \frac{1}{16a^3F^2}\sum_{i=\pi,K,\eta}
                \alpha_B^i\overline{Y}(a^2x_i^2) \nonumber \\
           && +\frac{1}{16aF^2}\sum_{i=\pi,K,\eta}x_i^2
              \beta_B^iW_4(a^2x_i^2).
      \label{mB3}
\end{eqnarray}
Similar notation will be used for the $\pi N$ sigma term,
\begin{equation}\label{sigmaexpansion}
\sigma_{\pi N} = \sigma_{\pi N}^{(0)} + \sigma_{\pi N}^{(1)}
               + \sigma_{\pi N}^{(2)} + \sigma_{\pi N}^{(3)} 
               + {\rm ~higher~order},
\end{equation}
where
\begin{eqnarray}
\sigma_{\pi N}^{(0)} &=& \sigma_{\pi N}^{(1)} = 0, \\
\sigma_{\pi N}^{(2)} &=& -2x_\pi^2(2b_0+b_{\cal D}+b_{\cal F}),
\end{eqnarray}
and $\sigma_{\pi N}^{(3)}$ is obtained by subtracting $\sigma_{\pi N}^{(2)}$
from Eq.~(\ref{sigma}).
Recall that the parameters $b_0$, $b_{\cal D}$ and $b_{\cal F}$ are not
dimensionless; they contain an implicit suppression factor due to their 
position in ${\cal L}_{\rm MB}^{(2)}$.

The four baryon masses plus the $\pi N$ sigma term are a set of five
observables which depend on five parameters: $m_0$, $b_0$, $b_{\cal D}$,
$b_{\cal F}$ and ${\cal D}$.  The other axial coupling is obtained
from ${\cal F} = g_A - {\cal D} \approx 1.267 - {\cal D}$.
To determine discretization errors (i.e. finite-$a$ errors;
differences from continuum results),
one could hold the parameters fixed and
determine the $a$-dependence of each observable, or conversely
hold the observables
fixed and determine the $a$-dependence of each parameter.
In what follows, the experimentally-measured 
values of the five observables will be used to 
determine the five parameters at each lattice spacing.

All five observables are linear in $m_0$, $b_0$, $b_{\cal D}$ and
$b_{\cal F}$ when the lattice spacing vanishes, but the
first term on the right-hand side of Eq.~(\ref{mB3}) introduces quadratic
dependences for nonzero $a$.  This term serves as a reminder that the
$a \neq 0$ extension of the continuum HBChPT Lagrangian is not unique.
In fact, the term can be eliminated by adding
new $a$-dependent terms to the original (minimal) HBChPT Lagrangian, such as
\begin{equation}\label{notunique}
\Delta{\cal L}_{\rm MB}^{(3)} \propto
 a{\rm Tr}\left(\bar{B}_vB_v\right)\left({\rm Tr}\chi_+\right)^2 + \ldots,
\end{equation}
with coefficients fixed appropriately.
In the present discussion, we will omit the first term on the right-hand 
side of Eq.~(\ref{mB3}) from calculations.  The particular discretization
errors obtained are expected to be representative of a typical lattice ChPT
Lagrangian.

Without the first term on the right-hand side of Eq.~(\ref{mB3}),
$\cal D$ and $\cal F$ are easily obtained from the Gell-Mann--Okubo
relation,
\begin{eqnarray}
\Delta_{\rm GMO}&=&\frac{3}{4}m_\Lambda + \frac{1}{4}m_\Sigma - \frac{1}{2}m_N
   - \frac{1}{2}m_\Xi \nonumber \\
  &=&\frac{({\cal D}^2-3{\cal F}^2)}{64a^3F^2}\left\{
     4\overline{Y}(a^2x_K^2)-\overline{Y}(a^2x_\pi^2)
    -3\overline{Y}(a^2x_\eta^2)\right\}.
\end{eqnarray}
With these couplings in hand, $b_{\cal D}$ can be extracted from
$m_\Sigma - m_\Lambda$, and $b_{\cal F}$ from $m_\Xi - m_N$.
Finally, $b_0$ is obtained from $\sigma_{\pi{N}}$ and $m_0$ from $m_N$.

\begin{figure}[tbh]
\epsfxsize=380pt \epsfbox[30 30 498 732]{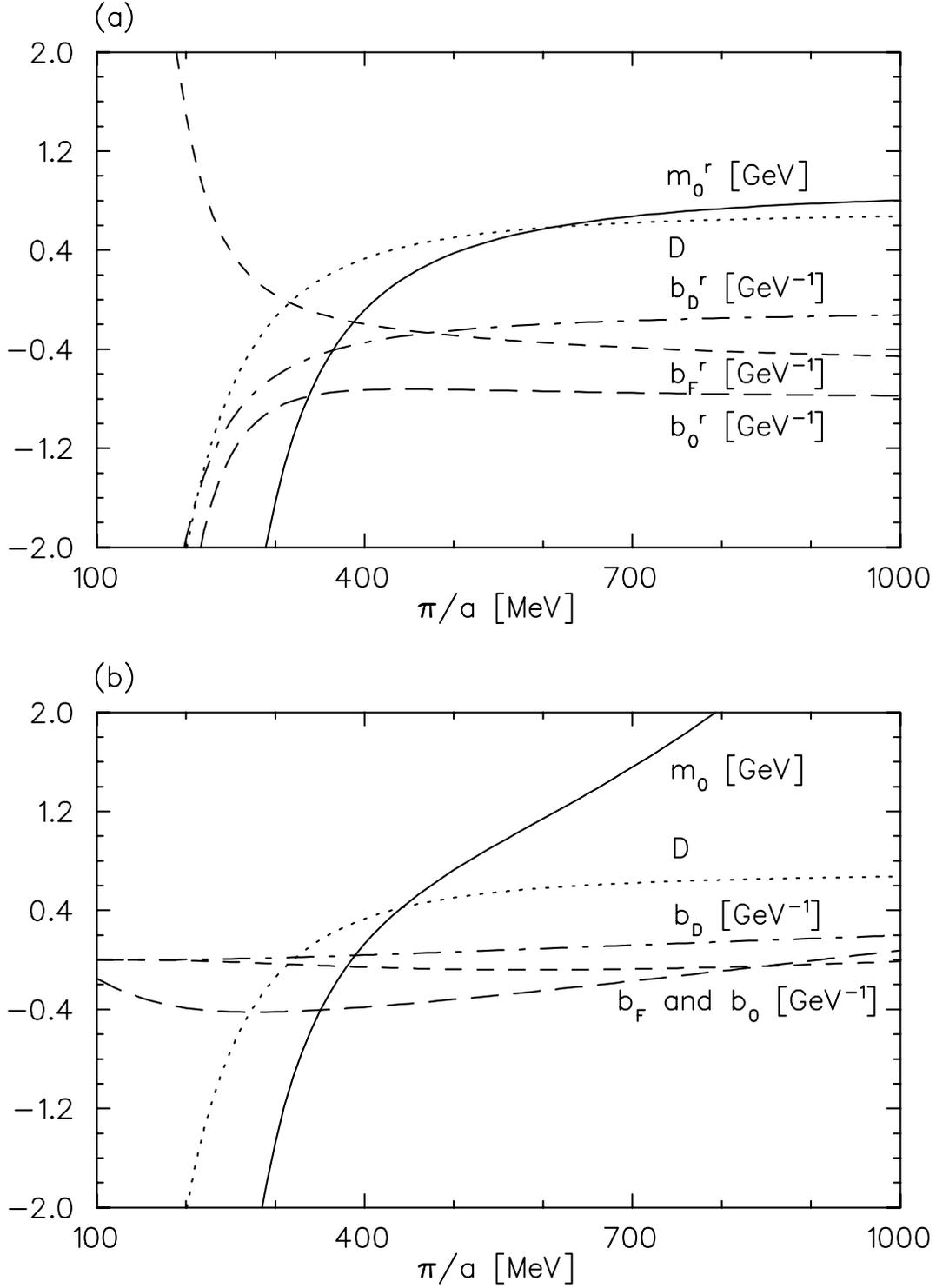}
\caption{Five parameters, obtained by fitting to the experimental values of
         $g_A$, $\sigma_{\pi{N}}$ and the four masses of the octet baryons.
         The sixth parameter is easily obtained as ${\cal F} = g_A - {\cal D}$.
         The fit to experimental values is redone for each lattice 
         cutoff, $\pi/a$.  Both the (a) renormalized and (b) unrenormalized
         values of the parameters are plotted.
         }\label{fig:params}
\end{figure}

Figure~\ref{fig:params} shows the resulting value of each parameter as a
function of the lattice cutoff.  The ``experimental'' value of $\sigma_{\pi N}$
was set to 45 MeV\cite{sigpiN}, 
and $m_\eta$ was required to satisfy Eq.~(\ref{mesonGMO}).
As expected, the renormalized parameters are essentially
independent of lattice spacing for $\pi/a\,\,\lower 2pt 
\hbox{$\buildrel>\over{\scriptstyle{\sim}}$}\,\,\Lambda_\chi$; 
their values in this
region are near the dimensional regularized values, which are ultimately
attained as $a\rightarrow 0$.  Significant lattice spacing dependences occur
for $\pi/a\,\,\lower 2pt \hbox{$\buildrel<\over{\scriptstyle{\sim}}$}\,\,
500$ MeV.  The cubic dependence of the unrenormalized parameter
$m_0$ on the inverse lattice spacing is
clearly seen.

\section{Conclusions}\label{sec5}

Lattice regularization is a method for introducing an ultraviolet cutoff
directly into the chiral Lagrangian without destroying the Lagrangian's
chiral symmetry.  As the lattice spacing vanishes, lattice regularization
represents an alternative to dimensional regularization.  Unlike dimensional
regularization, the lattice theory displays power divergences explicitly.

In this work, 
a lattice ChPT Lagrangian for mesons and baryons has been constructed, and
hadron masses and the pion-nucleon sigma term
have been calculated to one-loop order.

Researchers have had occasion to apply a nonlattice ultraviolet cutoff to ChPT.
(See, for example, Ref.~\cite{DonHB}.)  However, the spacetime lattice is a
particularly convenient implementation, due in part to its explicit
preservation of chiral symmetry.  Another benefit is that the lattice
spacing appears directly in the Lagrangian, whereas a nonlattice cutoff would
typically be invoked for each Feynman diagram at the time of loop integration.

Lattice regularization is also relevant to lattice QCD simulations,
where ChPT is routinely used to extrapolate lattice QCD data as a function
of quark masses.  The present work demonstrates
how discretization errors arising from the difference between lattice ChPT
and continuum ChPT can be calculated for a particular lattice ChPT
Lagrangian.

\section*{Acknowledgments}

It is a pleasure to thank Bu\={g}ra Borasoy and Nader Mobed for helpful 
discussions and comments on the manuscript.
This work was supported in part by the Natural Sciences and Engineering 
Research Council of Canada.

\begin{appendix}
\section{Limits involving $W_n(\epsilon^2)$ and 
         $\overline{Y}(\epsilon^2)$}\label{app:W}

Collected in this appendix are some limits involving the functions,
$W_n(\epsilon^2)$ and $\overline{Y}(\epsilon^2) \equiv Y_3(\epsilon^2)
+ Y_4(\epsilon^2)$, defined by
\begin{eqnarray}
W_n(\epsilon^2) &\equiv& \int_0^\infty{\rm d}x\,I_0^n(x)\exp\left\{-x\left(n+
\frac{\epsilon^2}{2}\right)\right\} \nonumber \\
&=& \int_{-\pi}^\pi\frac{{\rm d}^n\theta}{(2\pi)^n}\left(n+\frac{
\epsilon^2}{2}-\sum_{\mu=1}^n\cos\theta_\mu\right)^{-1}, \\
Y_n(\epsilon^2) &=& 4\int_0^\infty{\rm d}x\,I_0^{n-1}(x)\left\{I_0(x)
        -I_0^{\prime\prime}(x)\right\}\exp\left\{-x\left(n+\frac{\epsilon^2}
        {2}\right)\right\} \nonumber \\
&=& \frac{4}{n}\int_{-\pi}^\pi\frac{{\rm d}^n\theta}{(2\pi)^n}\left(\sum_{\mu
    =1}^n\sin^2\theta_\mu\right)\left(n+\frac{
\epsilon^2}{2}-\sum_{\mu=1}^n\cos\theta_\mu\right)^{-1}. \label{Yn}
\end{eqnarray}
As $\epsilon$ vanishes, the functions of interest approach a finite limit,
\begin{eqnarray}
\lim_{\epsilon\rightarrow 0}W_3(\epsilon^2) &=& W_3(0) \approx 0.51, \\
\lim_{\epsilon\rightarrow 0}W_4(\epsilon^2) &=& W_4(0) \approx 0.31, \\
\lim_{\epsilon\rightarrow 0}\overline{Y}(\epsilon^2) &=& \overline{Y}(0) 
             \approx 1.43.
\end{eqnarray}
By making use of the asymptotic behaviour of the Bessel function,
\begin{equation}\label{limbessel}
I_0(x)\stackrel{x\rightarrow\infty}{\rightarrow}\frac{\exp(x)}{\sqrt{2\pi{x}}},
\end{equation}
the following useful limits are obtained,
\begin{eqnarray}
\lim_{\epsilon\rightarrow 0}\left\{\frac{W_3(\epsilon^2)-W_3(0)}{\epsilon}
\right\} &=& -\frac{1}{2\pi}, \\
\lim_{\epsilon\rightarrow 0}\left\{\frac{W_4(\epsilon^2)-W_4(0)}{\epsilon}
\right\} &=& 0, \\
\lim_{\epsilon\rightarrow 0}\left\{\frac{\overline{Y}(\epsilon^2)-
      \overline{Y}(0)}{\epsilon}\right\} &=& 0.
\end{eqnarray}
When comparing lattice regularized results to dimensional regularization, the
following relation is helpful,
\begin{eqnarray}
\lim_{\epsilon\rightarrow 0}\left\{\frac{W_4(\epsilon^2)-W_4(0)}{\epsilon^2}
\right\} &=& \lim_{\epsilon\rightarrow 0}\left\{\frac{\ln\epsilon}{4\pi^2}
\right\} + {\rm ~constant}.
\end{eqnarray}
To understand this connection to the logarithm, integrate
Eq.~(\ref{limbessel}) to produce the exponential integral,
\begin{equation}
Ei(-x) = -\int_1^\infty\,{\rm d}t\,\frac{\exp(-xt)}{t},
\end{equation}
and then notice that
\begin{equation}
\lim_{x\rightarrow 0}\left(\frac{Ei(-x)}{\ln(x)}\right) = 1.
\end{equation}
Finally, the calculation of the $\pi{N}$ sigma term makes use of the following
limits,
\begin{eqnarray}
\lim_{\epsilon\rightarrow 0}\left\{\epsilon{W}_3^\prime(\epsilon^2)\right\}
 &=& -\frac{1}{4\pi}, \\
\lim_{\epsilon\rightarrow 0}\left\{\epsilon{W}_4^\prime(\epsilon^2)\right\}
 &=& 0, \\
\lim_{\epsilon\rightarrow 0}\left\{\epsilon\overline{Y}^\prime(\epsilon^2)
     \right\} &=& 0,
\end{eqnarray}
and the identity
\begin{equation}
Y_n^\prime(\epsilon^2) = \frac{2}{n} - \left(2+\frac{\epsilon^2}{n}\right)
W_n(\epsilon^2).
\end{equation}

\section{A loop calculation}\label{app:loop}

\begin{figure}[tbh]
\epsfxsize=380pt \epsfbox[30 320 498 450]{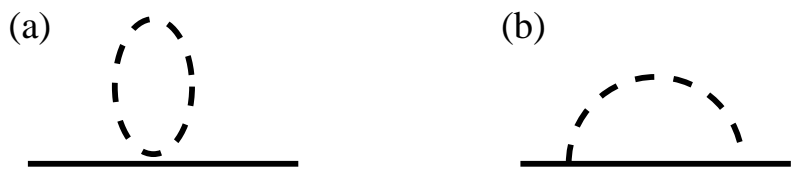}
\caption{The two loop topologies which contribute to a baryon mass
         at leading-loop order in HBChPT.
         }\label{fig:bloops}
\end{figure}

This appendix outlines the calculation of the Feynman diagram in 
Fig.~\ref{fig:bloops} which contains an intermediate baryon propagator.
For definiteness, the contribution of a charged-pion loop to the proton
is chosen.

The diagram represents the following integral,
\begin{eqnarray}\label{DeltaBBb}
\Delta\Gamma_{pp}^{(b)} &=& -ia\lim_{\epsilon\rightarrow 0}
\int_{-\pi/a}^{\pi/a}
\frac{{\rm d}^4p}{(2\pi)^4}\left\{x_\pi^2+\frac{4}{a^2}\sum_\nu\sin^2\left(
\frac{ap_\nu}{2}\right)\right\}^{-1} \nonumber \\
&\times& \left\{\sin(aq_4+ap_4+i\epsilon)
+2i\sin^2\left(\frac{aq_4+ap_4+i\epsilon}{2}\right)\right\}^{-1} \nonumber \\
&\times& \left\{\frac{i\sqrt{2}}{aF}({\cal D}+{\cal F})\sum_\mu{S}_\mu
     \sin(ap_\mu)\right\}^2.
\end{eqnarray}
The precise limits of integration deserve some thought.  One might consider
integrating from -min($\pi/a$,$\pi/a+q_\mu$) through min($\pi/a,\pi/a-q_\mu$)
since this would ensure that both the meson and nucleon momenta remain
within the first Brillouin zone.  However, choosing $v=(0,0,0,1)$ and 
working in the proton's rest frame makes $q_\mu$ suppressed by the inverse
baryon mass, so the limits of Eq.~(\ref{DeltaBBb}) are accurate to the 
order we are working.  These choices also lead to $S_4=0$.

Notice that an ``$i\epsilon$'' term has been included in the nucleon
denominator of Eq.~(\ref{DeltaBBb}), so that the singularity can be treated
correctly.
Using
\begin{equation}
\sin(aq_4+i\epsilon)+2i\sin^2\left(\frac{aq_4+i\epsilon}{2}\right)
= e^{-\epsilon}\{\sin(aq_4)+i[e^\epsilon-\cos(aq_4)]\},
\end{equation}
and a property of the Pauli-Lubanski vector,
\begin{equation}
2S_\mu{S}_\nu = \frac{1}{2}(v_\mu{v}_\nu - \delta_{\mu\nu}) + [S_\mu,S_\nu],
\end{equation}
one arrives at the following form,
\begin{eqnarray}
\Delta\Gamma_{pp}^{(b)} &=& \lim_{\epsilon\rightarrow 0}
\left(\frac{-({\cal D}+{\cal F})^2}{4aF^2}\right)\int_{-\pi/a}^{\pi/a}
\frac{{\rm d}^4p}{(2\pi)^4}\left\{x_\pi^2+\frac{4}{a^2}\sum_\nu\sin^2\left(
\frac{ap_\nu}{2}\right)\right\}^{-1} \nonumber \\
&\times& \sum_{k=1}^3\sin^2(ap_k)\left\{1+\left(\frac{\sinh\epsilon}
         {\cosh\epsilon-\cos(ap_4)}\right)\right\}.
\end{eqnarray}
The piece without $\sinh\epsilon$ is expressible in terms of $Y_4(a^2x_\pi^2)$,
and the piece containing $\sinh\epsilon$ can be expressed in terms of
$Y_3(a^2x_\pi^2)$.  The functions $Y_n(\epsilon^2)$ are discussed in
Appendix \ref{app:W}.
The final result is
\begin{equation}
\Delta\Gamma_{pp}^{(b)} = -\frac{3}{32a^3F^2}({\cal D}+{\cal F})^2
\left\{Y_3(a^2x_\pi^2)+Y_4(a^2x_\pi^2)\right\}.
\end{equation}

\end{appendix}

\newpage

\begin{table}[thb]
\caption{Coefficients that appear in the baryon masses.}\label{tab:coeffs}
\begin{tabular}{lllll}
 & $B=N$ & $B=\Sigma$ & $B=\Xi$ & $B=\Lambda$ \\
\hline
$\alpha_B^\pi$ & $(9/4)({\cal D}+{\cal F})^2$ 
               & ${\cal D}^2+6{\cal F}^2$ 
               & $(9/4)({\cal D}-{\cal F})^2$ 
               & $3{\cal D}^2$ \\
$\alpha_B^K$ & $(1/2)(5{\cal D}^2-6{\cal DF}+9{\cal F}^2)$ 
             & $3({\cal D}^2+{\cal F}^2)$ 
             & $(1/2)(5{\cal D}^2+6{\cal DF}+9{\cal F}^2)$
             & ${\cal D}^2+9{\cal F}^2$ \\
$\alpha_B^\eta$ & $(1/4)({\cal D}-3{\cal F})^2$ 
                & ${\cal D}^2$ 
                & $(1/4)({\cal D}+3{\cal F})^2$
                & ${\cal D}^2$ \\
$\beta_B^\pi$ & 3/4 & 3/2 & 3/4 & 1/2 \\
$\beta_B^K$ & 3/2 & 1 & 3/2 & 5/3 \\
$\beta_B^\eta$ & 5/12 & 1/6 & 5/12 & 1/2 \\
$\gamma_B^{\cal D}$ & $-4x_K^2$ & $-4x_\pi^2$ & $-4x_K^2$ & $-4x_\eta^2$ \\
$\gamma_B^{\cal F}$ & $4(x_K^2-x_\pi^2)$ & 0 & $-4(x_K^2-x_\pi^2)$ & 0
\end{tabular}
\vspace*{-5mm}
\end{table}


\end{document}